\documentclass[10pt,journal,compsoc]{IEEEtran}

\ifCLASSOPTIONcompsoc
  \usepackage[nocompress]{cite}
\else
  \usepack
\fi

\usepackage[pdftex]{graphicx}
\graphicspath{{../figs/}}
\DeclareGraphicsExtensions{.pdf,.jpeg,.png}

\usepackage{amsmath}
\usepackage{array}
\usepackage{mdwmath}
\usepackage{mdwtab}

\usepackage{fixltx2e}
\usepackage{stfloats}
\usepackage{url}

\usepackage{textcomp}                  
\usepackage{mathptmx}                  
\usepackage{cite}                      

\usepackage[font=footnotesize]{subcaption}
\usepackage{gensymb}
\usepackage{amssymb}

\usepackage[font=footnotesize]{subcaption}

\newcommand\T{\rule{0pt}{2.6ex}}       
\newcommand\B{\rule[-1.2ex]{0pt}{0pt}} 

\usepackage{color, colortbl}
\definecolor{LightCyan}{rgb}{0.88,1,1}

\begin{document}

\title{On a structural similarity index approach for floating-point data}

\author{Allison~H.~Baker, Alexander~Pinard, and~Dorit~M.~Hammerling
\IEEEcompsocitemizethanks{\IEEEcompsocthanksitem A.H. Baker is with the National Center for Atmospheric Research.\protect\\
E-mail: abaker@ucar.edu
\IEEEcompsocthanksitem A. Pinard and D.M. Hammerling are with Colorado School of Mines.\\
E-mail: apinard@ucar.edu, hammerling@mines.edu}
}

\maketitle


\begin{abstract}
Data visualization is typically a critical component of post-processing analysis workflows for floating-point output data from large simulation codes, such as global climate models.
For example, images are often created from the raw data as a means for evaluation against a reference dataset or image. 
While the popular Structural Similarity Index (SSIM) is a useful tool for such image comparisons, generating large numbers of images can be costly when simulation data volumes are substantial. In fact, 
computational cost considerations motivated our development of an alternative to the SSIM, which we refer to as the Data SSIM (DSSIM).  The DSSIM can be applied directly to the floating-point data to provide a general measure of whether images created from the data will be similar.
We evaluate the DSSIM in terms of quantifying differences due to lossy compression on large volumes of simulation data from a popular climate model. Bypassing image creation  results in a performance gain of two orders of magnitude for this case study. In addition, we show that the DSSIM is useful in terms of avoiding plot-specific (but data-independent) choices that can affect the SSIM.
While our work is motivated by and evaluated with climate model output data, the DSSIM may prove useful for other applications involving large volumes of simulation data.
\end{abstract}



\section{Introduction}\label{sec:intro}

Given the advances in high-performance computing (HPC) in recent years, model simulation codes are advancing scientific discovery across many disciplines. Visualizations of simulation output data are important to domain scientists as they typically provide the primary means of exploring the output data.
A typical post-processing data workflow often includes evaluating the differences between an image generated from the data and a reference image. Image quality assessment (IQA) measures are designed for this use-case, particularly the so-called ``full reference" (FR) IQA measures that require the original (or reference) image for the comparison (e.g., \cite{wang2011, fullref}).
The two images may differ due to differences in the datasets from which they were generated:
model simulation data versus experimental or observational data; two simulation datasets generated with slightly different model parameters or initial conditions; or the increasingly-common scenario where one image's data have been subjected to lossy compression.  
Regardless of the source of the discrepancy between images, IQA measures provide an objective way of quantifying that difference.  Perhaps the most well-known and commonly used IQA measure is the Structural Similarity Index (SSIM) \cite{wang04, Venkataramanan2021AHG}, though a number of such image-comparison measures have been developed (e.g., \cite{vif, gmsd, mad, fsim, Wang2009Mean}).

Because lossy compression remains a powerful tool in reducing the enormous volumes of simulation data produced with modern HPC machines, we are particularly interested in the use-case in which IQAs help to discover and quantify image artifacts due to lossy compression.  Recall that in contrast to lossless compression, applying lossy compression to a dataset prohibits exact reconstruction of the original data.
The SSIM is useful in this context as an objective means of assessing the effects of lossy compression.  For example, in the medical imaging field, images are typically compressed to reduce unmanageable data volumes, but clearly the potential loss of critical information, such as details needed to make an accurate diagnosis, is a concern.  Concordantly, the SSIM has been advocated as a means of evaluating compressed-medical-image quality in many studies (e.g., see \cite{g2012, wegener2010, Gaudeau2014, Razaak2013, Kowalik2014medImages}).
  
The application area of particular interest to us is climate modelling, where simulations are well-known for producing enormous amounts of output data (e.g., terabytes or even petabytes).  For a number of years, lossy data compression has been proposed as a means of mitigating the big data problem in climate research (e.g., \cite{hubbe2013, woodring2011, baker2014, kuhn2016}), though its acceptance in the climate community is far from secured as more comprehensive measurements for evaluating the loss of information are still needed.
Given the importance of data visualization to climate scientists interacting with model output, an objective means of assessing whether images generated from the compressed model data are noticeably different from images based on the original model data is critical.
Therefore, as part of an effort to persuade climate scientists to adopt lossy compression, we included the SSIM in a suite of measures to evaluate the ``quality" of compressed climate simulation data\cite{baker2017}. 
In a follow-up work \cite{baker2019}, we proposed a minimum threshold for SSIM values to indicate when differences could be seen when comparing images.  This threshold was based on a forced-choice visual evaluation study in which participants indicated whether a visual difference could be seen, with respect to the reference image that was created from uncompressed data. 
Note that evaluating the impact of lossy compression is a non-trivial task and depends on a number of factors such as the characteristics of the data, the type of compressor algorithm, and the intended scientific analysis of the data.  Therefore, multiple evaluation approaches are typically necessary to instill confidence in the compressed dataset.  Here, we do not analyze different methods of detecting compressor artifacts, as that has been done previously (e.g., \cite{baker2014, baker2017, poppick20}), but rather acknowledge that the SSIM is beneficial in this context and focus on how to reduce its cost.

While the SSIM is undoubtedly useful for objectively comparing images, several shortcomings arise in the context of its use in our compression-related research on large volumes of data. While we address those issues in this manuscript in the context of climate model data, the findings may be applicable to other application areas that rely on large volumes of floating-point simulation data. First, because the SSIM calculation is based on 
two corresponding grids of pixel values (i.e. a reference and a modified image), rendering images from the dataset values to be compared is required before the SSIM can be computed.
Generating many images, for example, from a long time series of climate data, potentially at high spatial resolution, can be quite computationally intensive. 
Indeed, the computational cost of generating the images from the climate model output required for the SSIM makes the SSIM a much more expensive measure of lossily compressed data quality than other data comparison measures that only require the floating-point data for identifying problematic compressor artifacts. 
However, we are loathe to abandon the SSIM due to cost considerations given its popularity and documented usefulness in image quality assessment \cite{Wang2009Mean} 
as well as our own positive experiences with it \cite{baker2019, baker2017}.

A second, albeit more minor, motivation is that the SSIM value is naturally dependent on plot parameters such as color scheme or geometric transformation (or other decisions that are not based on information contained in the data) that a scientist may make when creating a plot.  
As a result, for a particular floating-point dataset, two images that are considered indistinguishable based on their SSIM value given one set of plot parameters, may become distinguishable (again, based on the SSIM) if the plot parameters are changed.   A simple example of a plot setting for climate data that affects the SSIM is the global map projection. 
Therefore, because for our use-case we often do not know how an image will be generated from data, we would like the similarity measure that indicates whether an image created from a compressed dataset is likely to be distinguishable from that generated from the original dataset, to be independent of plot parameters.


Hence, the cost of generating images from the raw data 
as well as the possible dependence on plotting choices  
motivates applying the SSIM directly to the climate model's floating-point dataset values, rather than to the pixel values.
Unfortunately, simply applying the standard SSIM formula without modification to floating-point data, rather than pixel values of images created from the data, does not always result in desired behavior.
However, by making a few relatively simple but critical modifications, which we collectively refer to as the Data SSIM (DSSIM), we obtain a useful measure to apply directly to floating-point datasets 
that is comparable to the SSIM and better discriminates between the differences in our test datasets.
In this paper, we make the following contributions:
\begin{itemize}
    \item demonstrate the effect of basic data modifications and image generation choices on the computed SSIM value to improve our understanding of SSIM value ranges and dependencies;
    \item present a SSIM-like statistic that can be applied directly to floating-point data, thus avoiding the computational expense of rendering otherwise unnecessary  images;
    \item and provide an in-depth evaluation that illustrates the method's utility in evaluating the effects of  data compression on large volumes of climate data, particularly in terms of cost reduction.
\end{itemize}

The remainder of this paper is organized as follows.  In Section \ref{sec:back}, we review the SSIM and demonstrate its dependence on plot and parameter choices. Next, in Section \ref{sec:dssim}, we discuss considerations for floating-point data and introduce the DSSIM approach. 
In Section \ref{sec:app}, we discuss the applicability of the DSSIM in the context of evaluating lossy compression on climate data.  We provide concluding remarks in Section \ref{sec:concl}.

\section{Structural Similarity Index (SSIM)}\label{sec:back}

As previously noted, full reference IQAs are a popular means for comparing two images, where one image is typically the reference image, against which the quality of the second image is being compared.  The IQA value is intended to be an objective measurement of the more subjective concept of how noticeably different the two images are, say to a human observer. While the SSIM was developed to compare the encoding of natural images, we found in previous work \cite{baker2019} that the SSIM 
showed good predictive ability to gauge when experts perceive differences in images generated from climate model simulation data. In fact, while a number of other IQAs
showed good predictive ability, the SSIM IQA measure performed the best.   It is important to note that the plots of most interest to the  climate community in  diagnostic packages, for example, are typically those that smoothly map the floating-point data to RGB values.   In other words, pseudocolor plots (or possibly a filled contour - depending on the transfer function) that use a smooth colormap are suitable for comparison with the SSIM.  We do not consider scatter plots, data plots with a lot of white space, glyph-based techniques, or, more generally, plots for which a small change in the data causes a large or abrupt change in the image.  Also note that the SSIM can be sensitive to accessories on the plot such as plot grids, labels, etc. \cite{Veras20}, which should be removed for comparison purposes.

\begin{figure}[t]
\centering
\includegraphics[width=7.0cm]{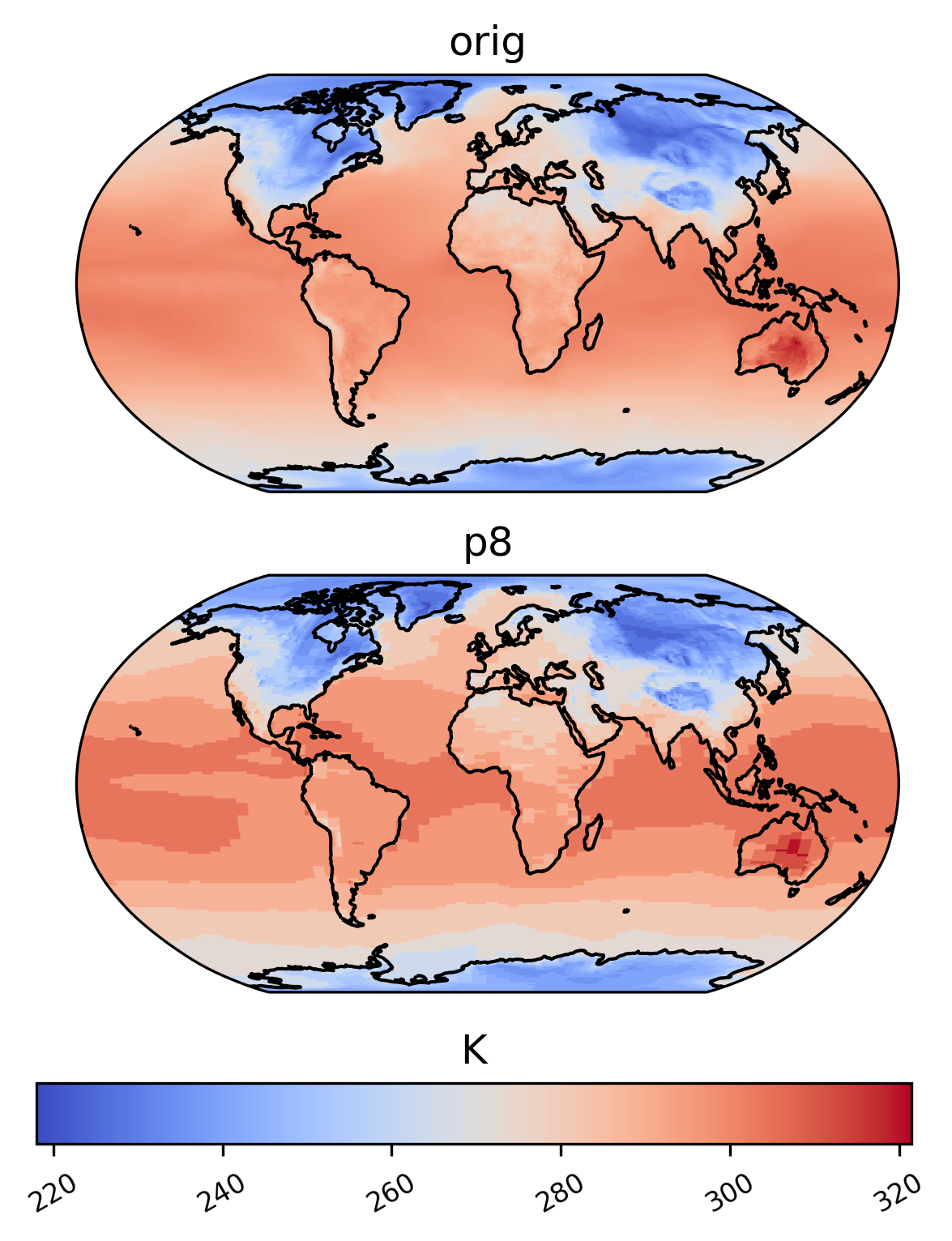} 
\caption{The plots for surface temperature (TS) data for the original data (top) and the lossily compressed data (bottom) with compressor \textsc{zfp} and p=8. (Compressor details are discussed in Section \ref{sec:exp}.) 
}
\label{f:ts_p8}
\end{figure}

Figure \ref{f:ts_p8} contains an example of the type of pseudocolor plots that climate scientists typically create for a commonly used climate variable.  Both plots in the figure show a single time slice of surface temperature (TS) data; the top plot contains the original (not compressed) data and the bottom plot contains data that has been aggressively (lossily) compressed such that artifacts are clearly visible.  The application dataset from which we obtained these data and the compressor are described in Section \ref{sec:app}.

 \subsection{Method overview}

The SSIM enjoys widespread use across a number of disciplines.  While some recent works cast doubt on the popular notion that the SSIM truly represents human visual perception (e.g., \cite{Dosselmann2011, nilsson2020understanding}), it nevertheless remains very popular in practice, due in large part to its simplicity 
and usefulness as a  statistical measure \cite{Venkataramanan2021AHG}.
The SSIM is the product of three factors that are intended to represent luminance, contrast,
and structure.  
Consider comparing two 
2D images $\mathbf{X}$ and $\mathbf{Y}$, each of dimension $m_x \times m_y$ with $M=m_x m_y$ pixel values.  The SSIM is computed by first calculating so-called per-pixel SSIM values comparing local patches (or windows) of the images. Let $\mathbf{x}_i$ and $\mathbf{y}_i$ be local image patches (i.e., 2D arrays that have been flattened) taken from the same location in $\mathbf{X}$ and $\mathbf{Y}$, respectively. Subscript 
$i$ indicates the pixel index in $\mathbf{X}$ and $\mathbf{Y}$ that is at the center of the local window ($i \leq M$).  And, let $N = n^2$ be the number of pixels in the local window.  Then, in the local window centered at pixel $i$ with $\mathbf{x}_i$ and $\mathbf{y}_i$ containing 
$N$ pixel values,  the per-pixel SSIM value is
\begin{equation}
SSIM(\mathbf{x}_i,\mathbf{y}_i) = (l(\mathbf{x}_i,\mathbf{y}_i))^\alpha \cdot (c(\mathbf{x}_i,\mathbf{y}_i))^\beta \cdot (s(\mathbf{x}_i,\mathbf{y}_i))^\gamma, 
\label{e:ssim_p}
\end{equation} 
where $l(\mathbf{x}_i,\mathbf{y}_i)$ is the luminance term, $c(\mathbf{x}_i,\mathbf{y}_i)$ is the contrast term, and $s(\mathbf{x}_i,\mathbf{y}_i)$ is the structure term. Parameters $\alpha$, $\beta$, and $\gamma$ are for adjusting the relative importance of the three terms).
The luminance, contrast, and structure terms for each local patch ${i}$ use the means ($\mu_{\mathbf{x_i}}$, $\mu_{\mathbf{y_i}}$), variances ($\sigma_{\mathbf{x_i}}$, $\sigma_{\mathbf{y_i}}$) and covariance ($\sigma_{\mathbf{x_iy_i}}$) that are computed on the local window, typically with Gaussian weights, from the arrays $\mathbf{x}_i$ and $\mathbf{y}_i$:
\begin{equation}
    l(\mathbf{x}_i, \mathbf{y}_i) = \frac{2 \mu_{\mathbf{x}_i} \mu_{\mathbf{y}_i} + C_1}
    {\mu^2_{\mathbf{x}_i} + \mu^2_{\mathbf{y}_i} + C_1},
\end{equation}
\begin{equation}
    c(\mathbf{x}_i, \mathbf{y}_i) = \frac{2 \sigma_{\mathbf{x}_i} \sigma_{\mathbf{y}_i} + C_2}
   {\sigma^2_{\mathbf{x}_i} + \sigma^2_{\mathbf{y}_i} + C_2},
\end{equation}
\begin{equation}
    s(\mathbf{x}_i, \mathbf{y}_i) = \frac{\sigma_{\mathbf{x}_i\mathbf{y}_i} + C_3}
    {\sigma_{\mathbf{x}_i} \sigma_{\mathbf{y}_i} + C_3}.
\end{equation}
Constants $C_1$, $C_2$ and $C_3$ are chosen to provide numerical stability by avoiding a zero denominator.
Note that the choice $C_1 = C_2 = C_3 = 0$ is equivalent to the universal quality index \cite{uiq} or UQI, which is a precursor to
the SSIM.
To simplify (\ref{e:ssim_p}), the following assumptions are suggested in \cite{wang04}: $\alpha = \beta = \gamma = 1$ and $C_3 = C_2/2$. This yields a simpler form of the per-pixel equation (\ref{e:ssim_p}):
\begin{equation}
  SSIM(\mathbf{x}_i,\mathbf{y}_i) =  S_1(\mathbf{x}_i,\mathbf{y}_i) S_2(\mathbf{x}_i,\mathbf{y}_i),
  \label{e:ssim_s}
\end{equation}
where
\begin{equation}
  S_1(\mathbf{x}_i,\mathbf{y}_j) =  \frac{(2 \mu_{\mathbf{x}_i} \mu_{\mathbf{y}_i} + C_1)}
    {(\mu^2_{\mathbf{x}_i} + \mu^2_{\mathbf{y}_i} + C_1)}
    \label{e:s1}
\end{equation}
and
\begin{equation}
  S_2(\mathbf{x}_i,\mathbf{y}_j) =  \frac{(2 \sigma_{\mathbf{x}_i\mathbf{y}_i} + C_2)}
    {(\sigma^2_{\mathbf{x}_i} + \sigma^2_{\mathbf{y}_i} + C_2 )}.
    \label{e:s2}
\end{equation}

Then the SSIM for the entire image, $SSIM$($\mathbf{X}$, $\mathbf{Y}$), which is sometimes referred to as the mean SSIM, is the average of the per-pixel SSIM values calculated for each local window ${i}$:
\begin{equation}
   SSIM(\mathbf{X}, \mathbf{Y}) = \frac{1}{M} \sum\limits_{i=1}^{M} SSIM(\mathbf{x}_i,\mathbf{y}_i).
   \label{e:ssim}
\end{equation}

The SSIM value has a couple of important properties.  First,  $SSIM(\mathbf{X}, \mathbf{Y}) = 1$ if and only if 
$\mathbf{X} = \mathbf{Y}$. Also,  $-1 \leq SSIM(\mathbf{X}, \mathbf{Y}) \leq 1$, and the closer $SSIM(\mathbf{X}, \mathbf{Y})$ is to 1, the more similar the images are. 
In practice, most calculated SSIM values are positive, with a negative value only occurring when the covariance term is negative (assuming nonnegative pixel values).

\subsection{Implementation}

For the implementation proposed in \cite{wang04}, the authors suggest the following constant values:
\begin{equation}
C_1 = (K_1L)^2 \qquad 
C_2 = (K_2L)^2,
\label{e:const}
\end{equation}
where $K_1 = 0.01$,  $K_2 = 0.03$, and $L$ is the dynamic range of the pixels (so $L=255$ for 8-bit images or $L=1$ if the image range is $[0,1]$). 
The authors note that the choice of these constants is ``somewhat arbitrary," but claim that the SSIM is ``fairly insensitive" to their values \cite{wang04}. 
The constant values are of particular interest when applied directly to floating-point simulation data, as discussed in the next section.

Another implementation detail is the local window (or patch) for which the per-pixel SSIM value statistics (mean, variance and covariance) are computed in (\ref{e:ssim}).  
The recommendation in \cite{wang04} is an $N = 11 \times 11$ window with a Gaussian filter kernel. 
Note that for the windows centered on the pixels at the boundaries of the image (i.e., within five pixels of the edge for the  $11 \times 11$ kernel), the Gaussian filter requires special treatment to handle the missing values (i.e., outside the image boundary).  However, in  the implementation in \cite{wang04}, these per-pixel SSIM values from the edge regions are simply excluded in the averaged SSIM value in (\ref{e:ssim}), 
which could lead to reduced emphasis on pixels near the edges of an image.

Because the SSIM is quite popular, many implementations are available.  We  
use the Python implementation of the SSIM that is available via \textsc{scikit-image} \cite{scikit-image}.
This version closely follows that of the simplified SSIM from \cite{wang04}, given here in (\ref{e:ssim_s}), 
using the default suggested parameters for the constants $C_1$ and $C_2$ and the Gaussian kernel size $11 \times 11$. 
This implementation similarly ignores the border per-pixel SSIM values when computing the overall mean SSIM. 
Note that for this \textsc{scikit-image} version, 
one must specify ``gaussian\_weights=True" and ``use\_sample\_covariance=False" to match the implementation in \cite{wang04}.   

\subsection{SSIM value dependencies } \label{sec:var}



\begin{figure}[t]
\centering
{\includegraphics[width=7.0cm]{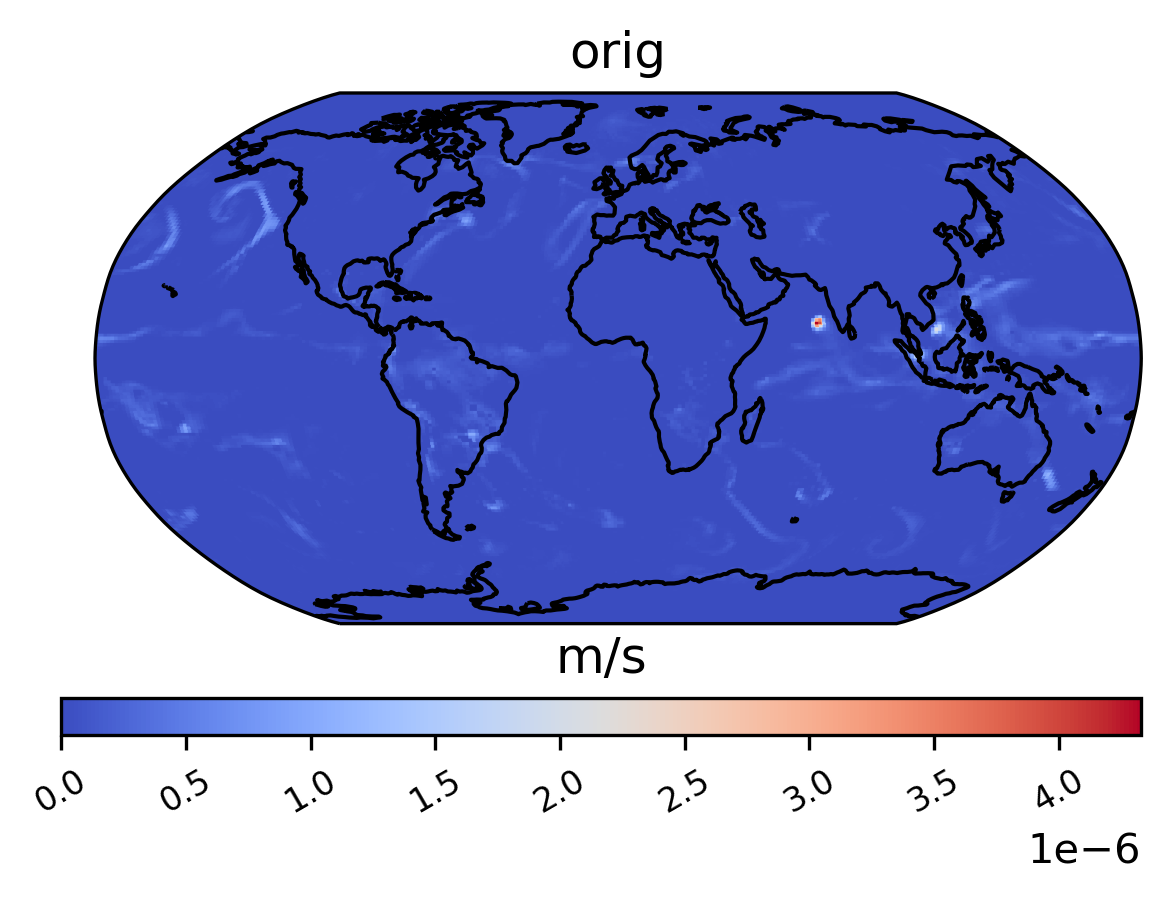} }
\caption{The default plot of original data  for precipitation rate (PRECT) in meters per second (m/s).
}
\label{f:prect_orig}
\end{figure}

The computed SSIM value for a pair of images naturally depends on a number of plot and SSIM parameter choices that are independent of the floating-point data.  Here we give a few examples to demonstrate the effect of these choices on the computed SSIM for two commonly used climate variables: surface temperature (TS) and precipitation rate (PRECT).   
Recall that the original TS data is shown in the top panel of Figure \ref{f:ts_p8}, and Figure
\ref{f:prect_orig} displays variable PRECT data.  These plots are representative of the types that climate scientists typically create and were generated 
with the \textsc{ldcpy}\footnote{\url{http://github.com/NCAR/ldcpy}} Python package,
which uses \textsc{Matplotlib} \cite{matplotlib} and \textsc{cartopy} \cite{Cartopy}.  
While our test data is further detailed in Section \ref{sec:app}, note that
these two sample variables have quite different characteristics: variable TS is relatively smooth and has a modest-sized range $\approx100 $K, while PRECT contains zero and near-zero values, changes more abruptly, and spans several orders of magnitude.

For reference, we briefly familiarize ourselves with the range of SSIM values that result from comparing climate variable images such as these. 
Table \ref{t:ts} lists the SSIM value in the third column that results from comparing the original TS data, as shown at the top of Figure \ref{f:ts_p8}, with several test data  cases (columns 4 and 5 are discussed in later sections).
For example, the LOSSY case uses the lossily compressed data shown in the bottom of Figure \ref{f:ts_p8}, and
the visual difference due to the rather aggressive choice of compression is quite obvious, resulting in an SSIM value of 0.93453. The remaining test data cases are generated as follows. 
We refer to the original TS data as $ts\_orig$ and its maximum and minimum values as $max$ and $min$, respectively. The data for the inverse (INV) case are computed by $inv\_data = max - ts\_orig + min$. 
Data for the RAND case are set to randomly generated values  between $min$ and $max$, and data for MEAN are constant values equal to the mean over all data in $ts\_orig$. Finally, the MIN data are all equal to $min$, ZERO indicates an array of all zeros, and PERT adds a random perturbation to $ts\_orig$ from $[1.0\text{e-}7, 0.1]$ Note that when comparing two images created from simulation data with the SSIM, one needs to ensure that both figures use the same colorbar, meaning that the transform from data to image space must be the same.  For the tests in Table \ref{t:ts}, we use the same colorbar shown in Figure \ref{f:ts_p8} with its extents set to the minimum and maximum values from the two datasets being compared.

The SSIM values for these test cases in column 3  of Table \ref{t:ts} may not be immediately intuitive.  For example, comparing to all zeros does not result in a SSIM value of zero.  
In fact, because the comparison uses the same colorbar for the transformation to image space,  the ZERO test case is more visually similar to the original than one would think.
In particular, because the colorbar range has been expanded to include zero, the 
original data visually appears much closer to constant-valued than in the top of Figure \ref{f:ts_p8}.   Thus, when compared to the constant of zeros values, the SSIM value is surprisingly far from zero (.72804).
Interestingly, the MIN and ZERO cases have nearly the same SSIM value, both of which
are noticeably lower than that of the MEAN case. 
Considering that the MEAN is also constant-valued, one might desire the SSIM to be lower than it is (.90384).  
Indeed, that the SSIM of MEAN is quite higher than that of the MIN is not necessarily intuitive, as
approximating the original data by a constant-valued array of the mean or minimum would both make for quite a poor compressor in terms of visual quality.  The RAND case is by far the worst according to the SSIM, which is rather expected as all structure is gone. The INV case is interesting as well as structure has been maintained in some sense, but the data are quite different.  These test cases are presented not to draw sweeping conclusions about the SSIM, but simply to illustrate what the SSIM value range may look like for a particular variable.

\begin{table}[tb]
\centering
\caption{A comparison of the original TS data to multiple test cases that are modifications of the TS test data.} \label{t:ts}
\footnotesize
\begin{tabular}{|l|l|l|l|l|l|}
\hline
case & data description & SSIM & SF-DSSIM & DSSIM \T \B \\
\hline \hline
LOSSY & ZFP with p=8  & .93453 & .84785 & .44112 \T \\
INV &  inverse values in [min,max] & .84872  & .28281 & -.70008 \\
RAND & random values in [min,max] & .34278 & .01050 & -.00061 \\     
MEAN & constant mean value & .90384 & .73068  & .00024 \\
MIN &  constant minimum value & .72282  & .70886 & .00000 \\
ZERO & all zero values & .72804 & .00013 & .00000  \\
PERT & add perturb. in [1.0e-7,0.1] & .99875 & .99995 & .98031 \B \\
\hline
\hline
\end{tabular}
\end{table}


\begin{table*}[t]
\centering
\caption{Examples of modifications to plot choices and constant values that affect the SSIM values.  Differences between the new and default SSIM values are given in parentheses (negative values indicate a decrease due to the modification).} \label{t:var}
\footnotesize
\begin{tabular}{|l|l|l|l|l|}
\hline
Setting description & Default & Modification & SSIM (TS)  & SSIM (PRECT) \T \\
\hline \hline
\rowcolor{LightCyan}
default & -  & -  & .99985  & .99153 \T \\
\hline
show coastlines & yes (width=0.5) & no & .99983 (-.00002) & .98958 (-.00195) \T \\
plot min/max   & TS:216.74/315.58 PRECT:-1.86e-8/1.18e-6 & TS:150/360 PRECT:-5.0e-8/5.0e-6 & .99995 (+.00010) & .99866 (+.00713) \\  
plot type & pcolormesh & contourf (TS:lev=25 PRECT:lev=50) & .99944 (-.00041) & .96864 (-.02289)  \\
figure dpi (dots/inch) & 300 & 100 & .99987 (+.00002) & .99323 (+.00170) \\
projection & Robinson & AlbersEqualArea     & .99857 (-.00128) & .99235 (+.00082) \\
projection & Robinson & PlateCarree & .99985 (.00000) & .99113 (-.00040) \\
colormap  & coolwarm & prism  &  .99544 (-.00441) & .88386 (-.10767)\\
colormap  & coolwarm & cool  &  .99994 (+.00009) & .99572 (+.00319) \B\\

\hline \hline
modified constants & K\textsubscript1=.01, K\textsubscript2=.03 & K\textsubscript1=.01, K\textsubscript2=.01 & .99909 (-.00076) & .96125 (-.03028) \T \\
modified constants & \textquotesingle\textquotesingle
 & K\textsubscript1=.01, K\textsubscript2=1e-8 & .99274 (-.00711) & .75449 (-.23704)\\
modified constants & \textquotesingle\textquotesingle & K\textsubscript1=.03, K\textsubscript2=.01 & .99909 (-.00076) & .96125 (-.03028)\\
modified constants & \textquotesingle\textquotesingle  & K\textsubscript1=1e-8, K\textsubscript2=.03 & .99985 (.00000) & .99153 (.00000) \\
modified constants & \textquotesingle\textquotesingle  & K\textsubscript1=1e-8, K\textsubscript2=1e-8 & .99274 (-.00711) & .75449 (-.23704)\\
modified constants & \textquotesingle\textquotesingle  & K\textsubscript1=1e-3, K\textsubscript2=1e-3 & .99298 (-.00687) & .82219 (-.16934) \\
modified constants & \textquotesingle\textquotesingle  & K\textsubscript1=0, K\textsubscript2=0 & NaN (--) & NaN (--) \B \\
\hline
\end{tabular}
\end{table*}

We now explore the effects of plot choices and SSIM constants on the SSIM calculation for both TS and PRECT and list the results in  
 Table \ref{t:var}.  Each row in the table compares the same two sets of data for both TS and PRECT.  
 In particular, the  TS and PRECT data used in this section are included in the \textsc{ldcpy} package in \texttt{data/cam-fv} directory (from NetCDF files \texttt{zfp1e-1.TS.100days.nc}, \texttt{orig.TS.100days.nc}, \texttt{zfp1e-7.PRECT.60days.nc}, and \texttt{orig.PRECT.60days.nc}).
The SSIM values in subsequent rows differ from the default due to either choices in the way the plots are generated or the SSIM is calculated; the underlying floating-point data is the same for each row. 
 The first row of the table, labeled ``default,"  lists the result of computing the SSIM via \textsc{ldcpy}, which uses the previously mentioned \textsc{scikit-image} SSIM implementation.  Recall  that the default images for the original TS and PRECT variable data are given in Figure 
\ref{f:ts_p8} (top) and Figure \ref{f:prect_orig}, respectively.   
The compressed data that we compare against are not shown as they have been chosen to be similar enough that differences cannot be seen for TS at this scale and are unlikely to be noticed for PRECT either, as indicated by the relatively high SSIM values for TS and PRECT: .99985 and .99153, respectively. We purposely chose a default  case with high SSIM values for each variable as values near a potential SSIM cutoff threshold (e.g., that indicates whether differences are noticeable to a human) are of most interest for our application. This concept of a cutoff threshold is discussed in the context of climate data in Section \ref{sec:app}.

\begin{figure}[t]
  \centering
  \subfloat[][Plot max/min: 150/360]{\label{f:maxmin}\includegraphics[width=4.1cm]{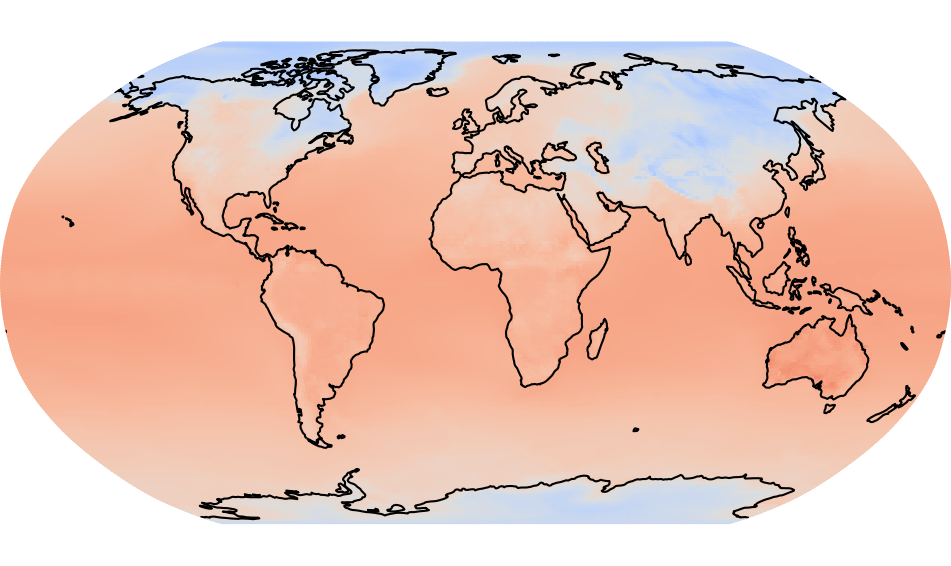}}\hspace{.1cm}
  \subfloat[][Plot type: contourf]{\label{f:ptype}\includegraphics[width=4.0cm]{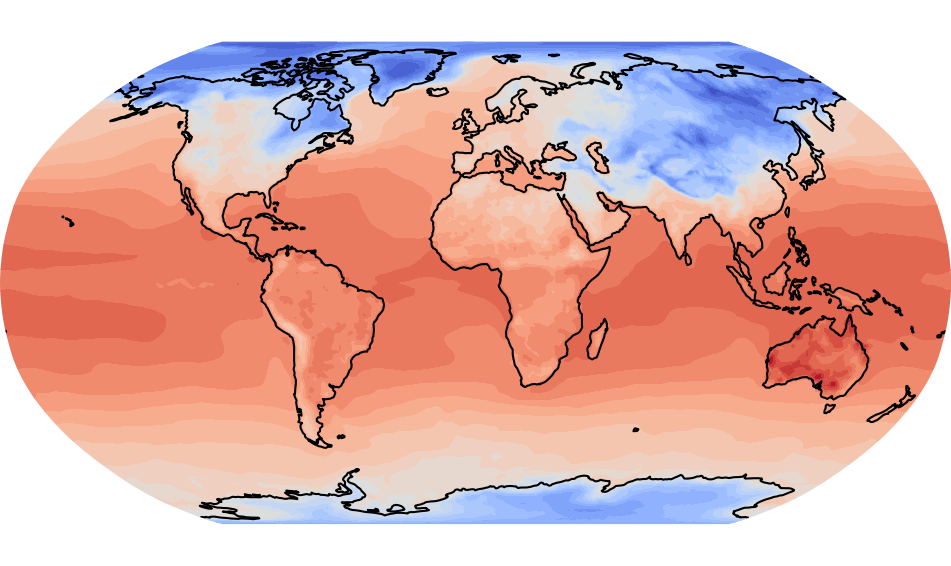}} \quad
  \subfloat[][Projection: AlbersEqualArea]{\label{f:proj}\includegraphics[width=4.1cm]{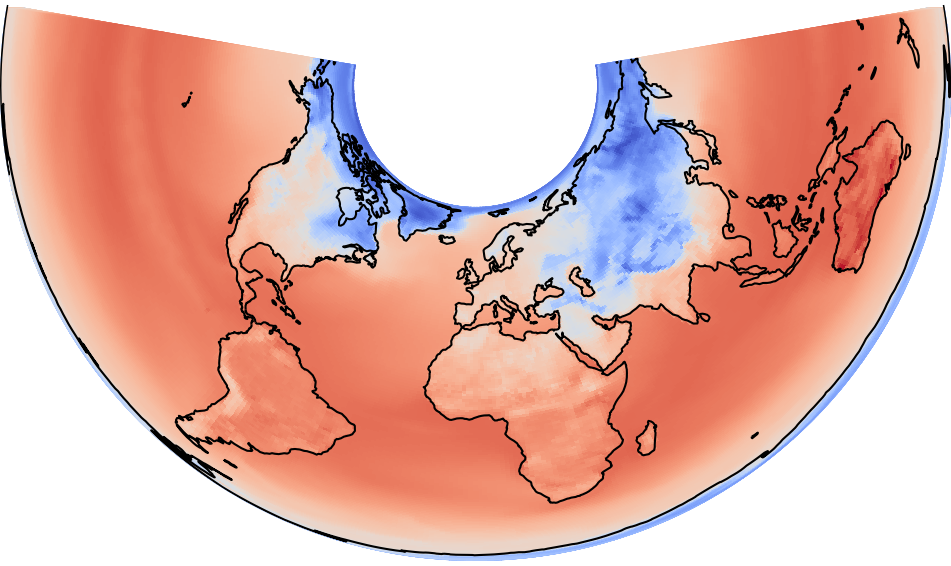}}\hspace{.1cm}
    \subfloat[][Projection: PlateCarree]{\label{f:proj-pc}\includegraphics[width=4.1cm]{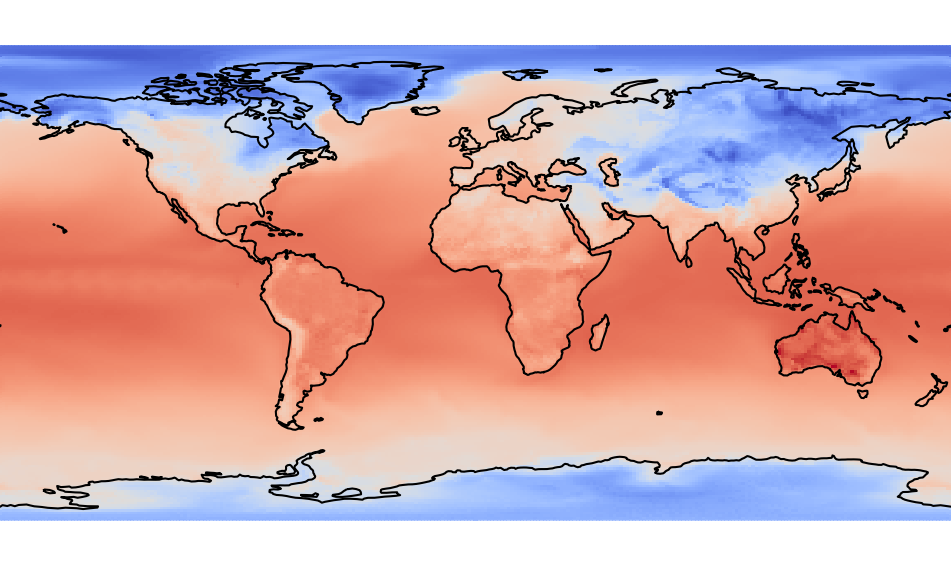}} \quad
  \subfloat[][Colormap: prism]{\label{f:color}\includegraphics[width=4.1cm]{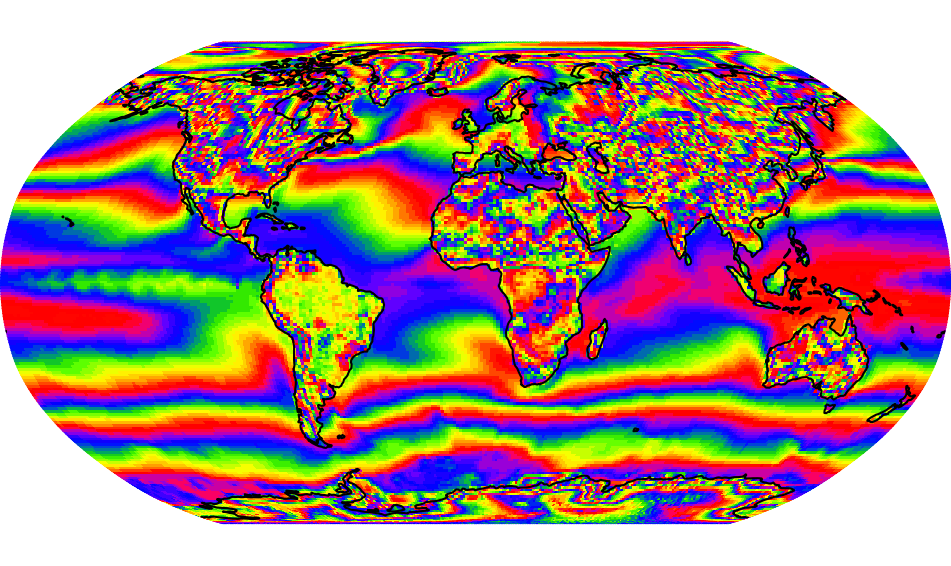}}\hspace{.1cm}
 \subfloat[][Colormap: cool]{\label{f:color-cool}\includegraphics[width=4.1cm]{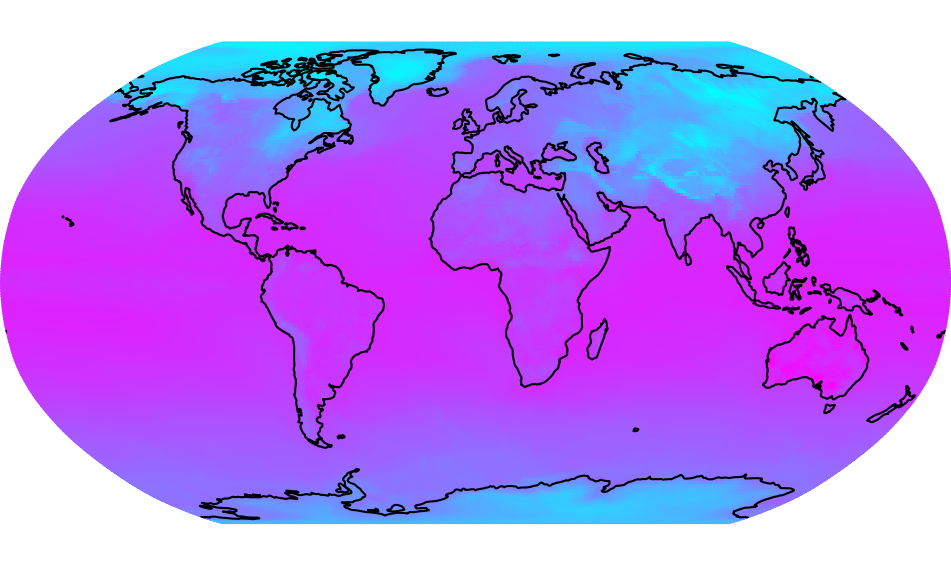}}\quad
  \caption{Examples of data-independent plot choices that can affect the SSIM values for surface temperature (TS).  Note that the corresponding SSIM values are given in Table \ref{t:var}.}
  \label{f:var}
\end{figure}

In the top half of Table \ref{t:var}, a subset of the parameter choices used to create these default plots are listed in the second column. 
For each row after the ``default" case, the parameter that is changed when creating both the original and compressed images is indicated in the third column, labeled ``modification".  
For reference, plots for a subset of these modifications for TS are displayed in Figure \ref{f:var}.
The differences in SSIM values give an idea of the effect of the plotting  choices for these two sample variables. Other climate variables may be  more or less sensitive to these changes, but that is not our focus.   
Note that while the first two significant digits are the same (.99) in each row for TS, we are interested in five significant digits to match the number of digits in the SSIM threshold from previously mentioned quality measures for climate compression \cite{baker2019}.
The second row shows that removing the coastlines does not have much effect on the SSIM for TS, but a bit more for PRECT.  
Enlarging the extents on the colormap (Figure \ref{f:maxmin}) as in row 3 of Table \ref{t:var} moves the SSIM closer to 1.0 for both variables, as would be expected. 
Figure \ref{f:ptype} shows the plot generated using \texttt{contourf()} instead of \texttt{pcolormesh()}, which also has a small effect for TS but a larger effect for PRECT as seen in Table \ref{t:var} (row 3).
The type of data projection onto a map, which is common for climate data, influences the SSIM as well.
The equal-area map projection (row 6) actually has more of an influence for TS (Figure \ref{f:proj}) than for PRECT. On the other hand, the often used equirectangular projection (row 7) has no effect on TS (Figure \ref{f:proj-pc}), but does affect PRECT.
The last two rows in the top half of the table illustrate that using a different colormap can definitely affect the SSIM, particularly when it is quite different from the original, such as that shown for TS in Figure \ref{f:color}.  The colormap changes the SSIM less (but still notably for PRECT) when it is similar to the original (Figure \ref{f:color-cool}).  Note that while the SSIM is not influenced by the color or hue of an image \cite{Veras20, Venkataramanan2021AHG, Hassan2012StructuralSM}, when we encode the floating-point data into a colormap, the characteristics of the colormap can influence the SSIM \cite{Veras20}. For example, because the \textit{prism} color map is more segmented than the default, this affects the SSIM more \cite{Veras20} than the \textit{cool} colormap which is more similar to the default \textit{coolwarm} map.  Other factors that we have not yet mentioned that
have been shown to affect the SSIM  include the size of the local window, whether or not to use Gaussian weights, and how the edge pixels are treated \cite{JONES201667}. 

The bottom half of Table \ref{t:var} focuses on the SSIM constants $K_1$ and $K_2$ as defined in (\ref{e:const}).  While it is customary to use the recommended defaults proposed in \cite{wang04}, 
we have found the SSIM calculation to be sensitive to the values for $K_1$ and $K_2$ (which in theory should only contribute to numerical stability) in some cases.
We note that the SSIM's sensitivity to constants is shown in \cite{sawant} as well, particularly the sensitivity to $K_2$, which is what we found with our climate data. 
Recall that the first row in Table \ref{t:var} (labeled ``default") gives the SSIM values for the default values of $K_1$ and $K_2$. For the number of significant digits that we list, only the effects of changing $K_2$ are noticeable.
Indeed, Table \ref{t:var} shows that for the PRECT data, changing $K_2$ can cause quite a large difference, exceeding anything shown in the top half of the table by orders of magnitude.

For our use case of interest and likely others, the observation that the SSIM calculation as in (\ref{e:ssim_s}) is more sensitive to $K_2$ than $K_1$ meets expectations.  In practice, the SSIM is generally used to compare images that are supposed to be similar in some sense (e.g., in our case,  differing only in compressor-induced artifacts). Therefore, it is reasonable to assume that the means of the two images will be quite similar.  As a result, for the first term in the SSIM, $S_1$ as given in (\ref{e:s1}), if $\mu_{x_i}$ and  $\mu_{y_i}$ are nearly the same over the local window, then the value of $C_1$ (and therefore $K_1$) is unimportant, and $S_1$ will be nearly one. However, the situation is less clear for the second term, $S_2$ as given in (\ref{e:s2}), in the SSIM calculation.  The magnitude of the numerator and denominator statistics, $\sigma_{x_iy_i}$ and  $(\sigma^2_{x_i} + \sigma^2_{y_i})$, respectively, may be less stable than the mean over the $11 \times 11$ window, in which case the value of $C_2$ (and thus $K_2$) becomes more influential.  This influence is important as we now move to applying the SSIM directly to the floating-point data.

A few considerations that we do not specifically address are related to the grid that the floating-point data live on.   
When an image is created from gridded data, the image may have either more or fewer pixels than grid points - depending on the chosen image resolution and the data grid size.
In addition, we are assuming that we have structured grid data.  

\section{Application to floating-point data} \label{sec:dssim}

Recall that our primary interest in the SSIM is as a tool for evaluating the effects of lossy compression on climate simulation data. 
While we also use other metrics to evaluate compression quality, the SSIM has proven useful for quantifying visual differences in images created from climate data \cite{baker2019}.  
The primary advantage of and motivation for applying an SSIM-like  statistic directly to the floating-point data (rather than the image pixel values) is the reduction in computational cost associated with generating images, particularly when data volumes are large and the images created are not actually needed for any other purpose.
A second more minor advantage in operating directly  on the  floating-point data is that we can avoid plot-specific but data-independent decisions (color, scale, axes, grid transform, etc.), which may result in different SSIM values for images created from the same datasets.
To summarize, then, our goal is to determine whether we can apply the SSIM
to the raw simulation data and obtain an indication of whether
differences in the data are likely to impact a visual assessment,
without committing to the creation of a specific set of images
from the data.

Finally, in developing a modified SSIM for floating-point data, we were motivated to rethink the choices for the constants.
While the SSIM constants were introduced to prevent dividing by zero \cite{wang04}, we would prefer that they do not noticeably affect the SSIM values, as shown in the previous section for PRECT especially. In particular, if the SSIM statistics in the local window (i.e., mean, covariance, variance) are close to zero for the floating-point data, then the constants may have an out-sized effect. This characteristic in not uncommon when comparing to data with modest compression.  Further, it is known that SSIM values tend to saturate toward one, and we see this effect even when the data are quite different, as for TS in Table \ref{t:ts}.   By making the constants less influential, we can spread the range away from one.

In this section, we first describe applying the SSIM formula, without modification, directly to floating-point data and explain why further modifications were desired.  We then discuss the further modifications that collectively result in the DSSIM and how they are useful. We compare the SSIM variants on climate model data with compression in Section \ref{sec:app}.

\subsection{A straightforward approach (SF-DSSIM)}\label{sec:sf-dssim}

The straightforward approach to extending the SSIM to floating-point data is simply to  use the SSIM equation in (\ref{e:ssim}) to compare the 2D arrays with $M$ grid points, where arrays $\mathbf{x}_i$ and $\mathbf{y}_i$ now contain the floating-point values in the local window of size $N$, where typically $N = 11 \times 11$, and are centered at grid location $i$.
The constant definitions and defaults in (\ref{e:const}) remain the same, but now $L$ is the dynamic range of the floating point data. 
We refer to this variant as straightforward data SSIM, or SF-DSSIM.

There are a number of considerations when applying the SSIM formula directly to floating-point data instead of to pixel values.
First, the suggested SSIM values for $K_1$ and $K_2$ may not be appropriate for every dataset.
While  for the SSIM, the pixel range L is typically L=255 or L=1, for floating-point numbers, the range may be much larger. 
If the range is quite large, then 
the suggested values for $K_1$ and $K_2$ may result in constants that are too big compared to the data values at some locations. 
This situation is of concern as the constants are only meant to prevent division by zero.  
Another scenario is that in which 
 the dynamic range for a set of data could be L=1, with all values in $[0,1]$, including many very small near-zero values (e.g., of order $1\text{e-}20$) as happens for the previously mentioned PRECT variable. In other words, if the data range is small but the range of exponents is quite large, then the constants will dominate the SF-DSSIM computation.  This size mismatch results in SF-DSSIM values of one or near one, even when the two datasets are quite different. (Recall that the SSIM value is 1.0 only when the two images are identical.)

Two more minor considerations when dealing with floating-point simulation data include that a NaN (or fill value or missing value) can be encountered in the data, which is common for climate data.
In this case, we do not want such values to propagate to the entire local window when the Gaussian filter is applied, and the code must handle this situation.
Another point is that while pixel values are typically nonnegative, floating-point values are often negative (or a mix of positive and negative).  Therefore, the situation where the sign of means $\mu_{\mathbf{x_i}}$ and $\mu_{\mathbf{y_i}}$ are opposite can occur and cause
  $S_1(\mathbf{x_i},\mathbf{y_i})$ as given in (\ref{e:s1}) to be negative.

In Table \ref{t:ts}, the second column from the right lists the SF-DSSIM values
for the modified TS datasets.  Note that the MIN and ZERO cases have nearly the same SSIM values, but quite different SF-DSSIM values, which may arguably make more intuitive sense in terms of a comparison to all zeros resulting in a nearly zero SSIM value. For the PERT case, the SF-DSSIM is actually closer to 1.0 than the SSIM value. As will be discussed further in the context of climate data, this behavior is largely the result of the quantization step in rendering the images for the SSIM calculations.  This particular perturbation results in quantization bin changes (which can increase the difference) for some of the values perturbed at the high end of the interval. If the perturbation values were all small enough, e.g. in  $[1.0\text{e-}7, 1.0\text{e-}5]$, then the SSIM and SF-SSIM would both be 1.0 (to five significant digits).

\subsection{Data SSIM (DSSIM)}\label{sec:dssim-sub}

We now explain the modifications to the SF-DSSIM approach that collectively result in our proposed variant of the SSIM for floating-point data, which we refer to as DSSIM.
To begin, we normalize both sets (the original and that to compare) of floating-point data to the range $[0,1]$. We normalize the data for a couple of reasons.  First, normalizing to this range makes determining appropriate constants, which we discuss shortly, much easier.  This step 
both eliminates the need for the $L$ term in the constants in (\ref{e:const}), as $L = 1$, and ensures that $S_1(\mathbf{x_i},\mathbf{y_i})$ is nonnegative, as is typically the case with the SSIM (meaning that a negative value can only result from a negative covariance).  Also, when visualizing floating-point data, the first step is to transform the data to the color bar range. We assume a linear transform and note that pixel values are often normalized so that each pixel value has a value between 0 and 1.

\begin{figure}[t]
\centering
\includegraphics[width=7.0cm]{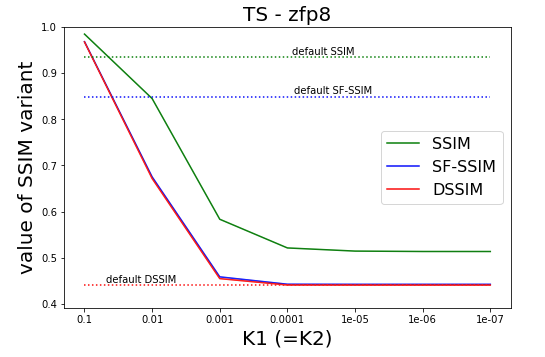} 
\caption{This plot shows the effect of modifying the value of $K_1 (=K2)$ when computing the SSIM, SF-DSSIM, and DSSIM for comparing the original
surface temperature (TS) data
to the lossily compressed data with compressor \textsc{zfp} and p=8.)
}
\label{f:ts_const}
\end{figure}

In determining the choice of constants for DSSIM, recall that
the two constants $C_1$ and $C_2$ are intended to provide numerical stability.  We simply want $C_1$ and $C_2$ to be small enough to not disproportionately influence the value of the DSSIM,
yet big enough to prevent dividing by zero.
We set them to equal values largely for convenience:  $C_1 = C_2$ and $K_1 = K_2$. Therefore, because $L=1$ for the DSSIM, we have $C_1 = K_1^2$, and 
we find that
\begin{equation}
C_1 = K_1^2  = 1\text{e-}8 ,
\end{equation}
is a reasonable choice for DSSIM.
We verified this choice for our application data by examining the influence of changing the constant on the DSSIM calculation for a number of different variables and compressor levels and finding the largest constant value that no longer influences the DSSIM value. 
For example, in Figure \ref{f:ts_const}, we show the effect of changing the constant values for  the SSIM, SF-DSSIM, and DSSIM when comparing the original TS data with the lossy compressed version shown in Figure \ref{f:ts_p8} and listed in the first row of Table \ref{t:ts}.  The dashed lines indicate the default value of the three SSIM approaches, which is at approximately $K_1 = K_2 = .01$ on the plot for SSIM and SF-DSSIM (``approximately"  because $K_2 = .03$ in these methods, which results in a subtle difference). The plot shows that the DSSIM value no longer decreases with decreasing constant at $K_1= .0001$, and this behavior is representative of what we observed for test data from other variables. While this choice is likely reasonable for data from other application areas as well, verifying that the computed DSSIM values are not sensitive to the constants is certainly easy to check.

Another difference for DSSIM is that after normalizing the floating-point data to the range $[0,1]$, DSSIM quantizes the data into 256 bins.  
The quantization mimics a linear color map transformation by allowing the DSSIM to use a similar precision on the floating-point data as the SSIM is using on image pixel data.  This step is particularly needed when comparing data that is quite visually similar, meaning that the SSIM value is quite close to one. We demonstrate and further explain the effects of this modification for climate variables in the next section.

Finally, we address any NaN values (or fill or missing values, which we hereafter also refer to as NaNs) present in the data when the Gaussian kernel is applied locally to each $11 \times 11$ window.  
If the center point of the window (grid point $i$) is NaN, that window calculation is simply excluded from the final mean SSIM calculation given in (\ref{e:ssim}).  
However, if the value at the center of the window is not NaN but any of the other local values in the window are NaN, the filter must be modified (otherwise the $DSSIM(\mathbf{x_i},\mathbf{y_i})$ value will be set to NaN). 
In the convolution and 2D Gaussian kernel functions in the \textsc{astropy}\footnote{\url{http://www.astropy.org}} package \cite{astropy2013} \cite{astropy2018}, NaNs are replaced by interpolating from neighboring data points within a given kernel.
In particular, for the kernel, DSSIM uses \texttt{Gaussian2DKernel(x\_stddev=1.5, x\_size=11, y\_size=11)} and then we convolve with \texttt{filter\_args = {'boundary': 'fill', 'preserve\_nan': True}}.
The boundary option for convolve is not important in the current implementation as the DSSIM values whose windows extend past the boundary are ignored when computing the mean DSSIM over the grid -- as with the SSIM.
Note that \textsc{scipy} convolution routines do not properly deal with NaNs at this time, but our
DSSIM implementation, available in the previously mentioned \textsc{ldcpy} package, does.  

Referring back to Table \ref{t:ts}, the rightmost column lists the DSSIM values.  Note that the negative value for the INV case helpfully reflects the negative correlation in the data.  The three constant-valued sets MEAN, MIN, and ZERO are now all similar and very close to zero.  RAND is now also near zero.  From our user base's point of view, the DSSIM values are more intuitively in line with the type of changes represented by the test cases.
In the next section, we will use climate simulation data to demonstrate the benefit of using the DSSIM.


\section{Application to climate data}\label{sec:app}

We now describe our investigation into whether we can use the DSSIM instead of the SSIM for evaluating the effects of lossy compression on climate data in order to avoid the computational cost incurred by generating images.
We use data from the
Community Earth System Model (CESM) \cite{cesm2013}, which is a popular climate model that generates far too much data (e.g., terabytes or petabytes) -- hence the interest in lossy compression. 
In previous work in  \cite{baker2017} and \cite{baker2019}, we applied the SSIM to images created from CESM data with the NCAR Command Language (NCL) \cite{ncl}, which are similar to those generated by the Atmosphere Working Group Diagnostics Package (AMWG-DP).  
While AMWG-DP-type images are familiar to scientists in the Earth science community because of its historical widespread use, Python has been quickly replacing NCL as the analysis tool of choice in recent years. In fact, many scientists are doing their own analyses in Python with the help of communities such as Pangeo\footnote{\url{http://pangeo.io}}\cite{pangeo} and creating their own images. This change in post-processing analysis provided further motivation for us to use an SSIM-like measurement that is independent of plot choices.

\subsection{Experimental data details} \label{sec:exp}

The experiments in this paper use a subset of data from the popular and publicly available CESM Large Ensemble Community Project (CESM-LENS) \cite{kay2015}.
In particular, we use the CESM-LENS data corresponding to the RCP8.5 forcing period, which begins in January 2006, and ensemble member 31.
We focus on the atmospheric model output, which uses a one-degree latitude-longitude grid corresponding to $192 \times 288$ grid points per vertical level and 30 vertical levels. 
Each of the more than 200 atmospheric variables is stored in a NetCDF-formatted time-series file according to its output frequency: monthly, daily, or 6-hourly.
And while CESM performs computations in double precision (64-bit), it writes data to file in single precision (32-bit). In Table \ref{t:var_props}, we list a subset of the atmospheric variables for which we show results. These variables are among the most frequently downloaded and analyzed from the CESM-LENS dataset and have differing characteristics. For example, in terms of compression, TS is considered an ``easy" variable to compress due to its relatively narrow range and smoothness,
while variables such as PRECT, which has a large range of values, including some very small values close to zero, are typically challenging for lossy compressors.


For these experiments, we compress CESM-LENS data with the popular \textsc{zfp} compressor
\cite{lindstrom2014}.
\textsc{zfp} is a high-speed lossy compressor 
designed for compressing logically regular and spatially correlated arrays of floating-point numbers, compressing data based on various accuracy or size constraints.
We use \textsc{zfp} 0.5.5 in fixed-precision mode, meaning that the precision encoded for the transform coefficients is fixed.  The fixed-precision mode parameter ($p$) specifies how many uncompressed bits per value to store (related to the relative error), so the smaller the value of $p$, the more aggressive the compression. To improve compression quality, we also use a newer \textsc{zfp} feature that addresses biased error 
and is available in \textsc{zfp} 0.5.5 by checking out the  ``feature/unbiased-error'' branch from the \textsc{zfp} Github page. In particular, we enable the pre-rounding mode by configuring \textsc{zfp} with 
``\texttt{cmake -DZFP\_ROUNDING\_MODE=ZFP\_ROUND\_FIRST -DZFP\_WITH\_TIGHT\_ERROR=ON}".

\begin{table*}[t]
\centering
\caption{Sample atmospheric variables  with their descriptions and selected characteristics (minimum, absolute nonzero minimum, and maximum values) for the first time slice. Note that the absolute nonzero minimum is not listed when it is equivalent to the minimum. All listed variables are 2D. (CLOUD is normally a 3D variable, but here we use vertical level 20 only.)} \label{t:var_props}
\footnotesize
\begin{tabular}{|l|l|l|l|l|l|l|l|}
\hline
variable &             &       & output   &        &     & abs. min &  \T \\
name     & description & units & frequency& mean   & min & (nonzero)  & max  \B \\
\hline \hline
TS  & surface temperature & $K$ & daily & 284.64 & 218.12 & -- & 321.47   \T \\
PRECT & precipitation rate & $m/s$ & daily & 2.74e-8 & -1.16e-20 & 8.60e-32 &  4.33e-6  \\
PS & surface pressure & $Pa$ &  monthly & 9.85e4 & 5.24e4 & -- & 1.04e5 \\
FLUT & upwelling long-wave flux (top of model) &  $W/m^2$ & daily & 233.72 & 80.529 & -- & 338.08 \\
QFLX & 	surface water flux & $kg/m^2/s$  & monthly & 3.53e-5 & -2.26e-6 & 1.57e-11 & 1.44e-4  \\
CLOUD & cloud fraction & fraction & monthly & 0.14 & 0.0 & 4.86e-12 & 0.74 \B \\
\hline
\hline
\end{tabular}
\end{table*}

\subsection{SSIM variants and compressed climate data}

\begin{figure}
\centering
\subfloat{\includegraphics[width=7cm]{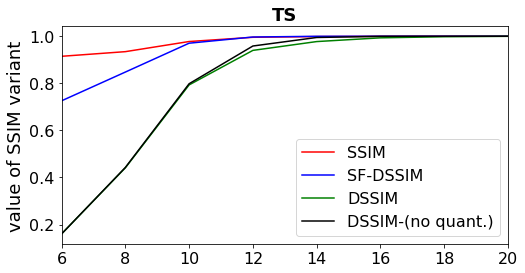} }
\newline
\subfloat{\includegraphics[width=7cm]{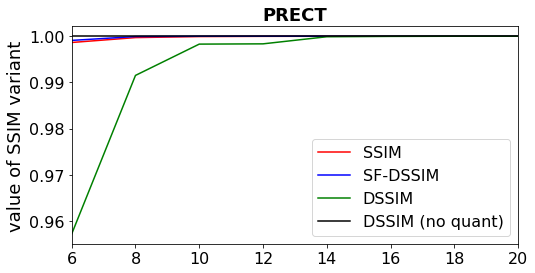} }
\newline
\subfloat{\includegraphics[width=7cm]{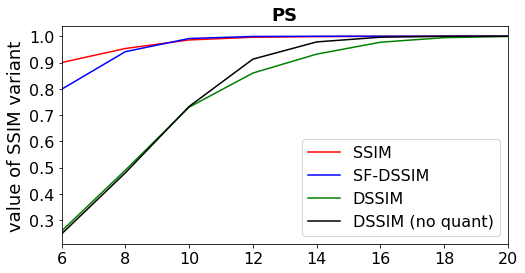} }
\newline
\subfloat{\includegraphics[width=7cm]{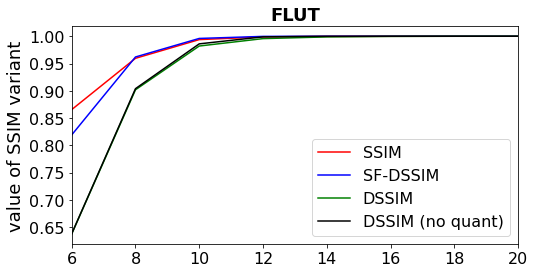} }
\newline
\subfloat{\includegraphics[width=7cm]{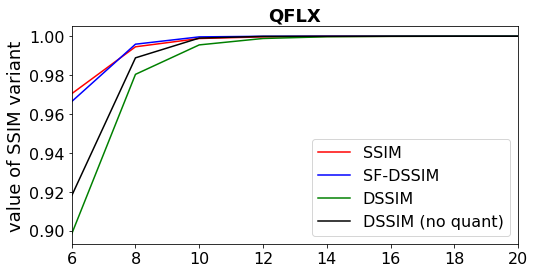} }
\newline
\subfloat{\includegraphics[width=7cm]{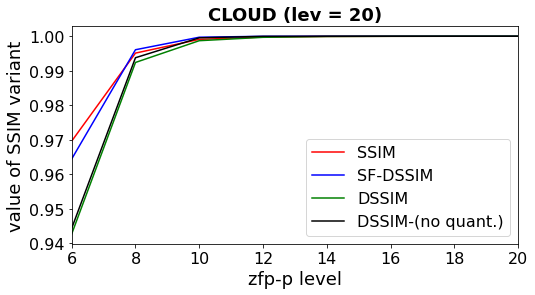} }
\newline
\caption{A similarity comparison between the original data and the  \textsc{zfp}-compressed data for different compression parameter ($p$-level on the x-axis).  The range on the y-axis differs for each plot.
}
\label{fig:all}
\end{figure}

To better illustrate their differences, we compare the SSIM, SF-SSIM, and DSSIM on the atmospheric variables in Table \ref{t:var_props}, each of which is compressed by varying amounts.
The SSIM quantities are again computed via \textsc{ldcpy} with default settings as in Section \ref{sec:var}, and results for each variable are given in Figure \ref{fig:all}.
For each, we plot the SSIM, DSSIM, and SF-DSSIM values that compare the original dataset to its compressed version.  We show results for eight different levels of \textsc{zfp} fixed-precision compression ($p = 6,8,10,12,14,16,18,20$).
As $p$ increases, the compressor is increasingly conservative. In particular, $p=6$ is the most aggressive compressor option shown and should result in the smallest SSIM values for all variables, and $p=20$ is fairly conservative and should result in values close or even equal to one. 
We do not show $p>20$ as they are indistinguishable from each other and 1.0 in the plots.
A fourth SSIM variant appears in these figures as well: ``DSSIM (no quant)". This approach is equivalent to DSSIM without the quantization step and will be discussed shortly.

Recall that the primary motivation for using the DSSIM instead of the SSIM is to avoid the cost associated with generating plots to compute the SSIM that are otherwise not needed, as is the case here for evaluating lossy compressor artifacts.  
In applying the SSIM formula directly to the floating-point data,
we took the opportunity to make a few beneficial modifications, collectively referred to as the DSSIM, whose effects can be seen in the
plots in Figure \ref{fig:all}.    
First, note that by normalizing the data  to $[0,1]$ and  then choosing constants that do not have an out-sized influence on the computed SSIM values (e.g., as discussed for Figure \ref{f:ts_const}), the DSSIM obtains much lower values for more aggressive compressor options, i.e., smaller values of $p$, than the SSIM and SF-DSSIM.  
The slope of the DSSIM line is much steeper in this aggressive-compression region. 
Furthermore, by mitigating sensitivity to the constants, 
the DSSIM behavior resulting from more appropriate constants is desirable to us as intuitively we want compressor choices, that are too aggressive, to have a more noticeable drop in the DSSIM value.

Another trend in Figure \ref{fig:all} is that as the compression becomes more conservative with increasing $p$ values, the SF-DSSIM values reach 1.0 sooner than the SSIM for all of the variables. 
Recall that for the SSIM, values of 1.0 are obtained only when the 
images are exactly the same.  When applying this calculation directly to the floating-point data, it is reasonable to not want a value of 1.0 when the SSIM value is still below 1.0, indicating that the images are not equivalent.  
Instead, 
the DSSIM's sensitivity to  small differences when the data is quite similar (higher values of $p$) can be beneficial for identifying minor differences.
Because identifying when each SSIM variant reaches 1.0 (to five significant digits) is difficult to discern in the Figure \ref{fig:all}, we list the corresponding \textsc{zfp} parameter $p$ at which this occurs in Table \ref{t:one}.
In our experience, the DSSIM value is always less than the SSIM value, whereas the  SF-DSSIM lines typically intersect the SSIM line at some point, because for smaller $p$, the SF-DSSIM is usually smaller than the SSIM (but then levels off to 1.0 more quickly).  This statement may not be universally true as the SSIM is sensitive to plot choices, but  holds for all the variables that we have examined in this representative climate dataset.
As mentioned previously, recall that we assume that the floating-point data is smoothly mapped to RGB values,  as with pseudocolor plots that use a smooth colormap.

\begin{table}[tb]
\centering
\caption{A list of the smallest \textsc{zfp} compression parameters ($p$) for which the SSIM variants equal 1.0. Parameters in red indicate that the SSIM variant (SF-DSSIM, DSSIM, or DSSIM (no quant.)) reached 1.0 for a more aggressive compression parameter (lower $p$) than the original SSIM value (bold) did, which is undesirable.} \label{t:one}
\footnotesize
\begin{tabular}{|l|l|l|l|l|l}
\hline
variable  & SSIM & SF-DSSIM & DSSIM & DSSIM (no quant.) \T \B \\
\hline \hline
TS  & \textbf{24} & \textcolor{red}{18} & 24 & \textcolor{red}{20} \T \\
PRECT  & \textbf{16} & \textcolor{red}{10} & 18 & \textcolor{red}{14}\\
PS & \textbf{22} & \textcolor{red}{18} & 24 & 22\\
FLUT & \textbf{20} & \textcolor{red}{16} &  22 & \textcolor{red}{16}\\
QFLX &  \textbf{20} & \textcolor{red}{14} & 22 & \textcolor{red}{14}\\
CLOUD &  \textbf{18} & \textcolor{red}{14} & 18 & \textcolor{red}{14} \B \\
\hline
\hline
\end{tabular}
\end{table}

Figure \ref{fig:all} and Table \ref{t:one} also include results from running the DSSIM without the quantization step, as it is more difficult to intuit the usefulness of this modification.  In fact, for TS and FLUT variables, the quantization has virtually no noticeable effect in the subplots in Figure \ref{fig:all}, and for the remainder of the variables, it is difficult to see what is happening when the SSIM value is approaching one.  
Again, Table \ref{t:one} better illustrates the situation by providing the DSSIM results without quantization in the rightmost column.  As with the SF-DSSIM, DSSIM (no quant.) reached 1.0 before (i.e., at a smaller/more aggressive value of $p$) the SSIM for 5 of the 6 variables.  
For TS and CLOUD, the SSIM and DSSIM reach 1.0 at the same compressed dataset parameter $p$, and the DSSIM reaches 1.0 under more conservative compression for the remainder.  This behavior is acceptable as we prefer to err on the conservative side, and these SSIM and DSSIM values near 1.0 are important when determining a similarity cutoff threshold, which we address in the next subsection.
And, as previously discussed, we do not want a value of 1.0 from a similarity measure on the floating-point data when the SSIM on the images is less than 1.0, implying that there is a visual difference. Note that the data compression here with \textsc{zfp} is lossy for all tested values of $p$, meaning that the two floating-point datasets being compared are not equivalent.  However, we do \textit{not} have the requirement that a DSSIM of 1.0 comparing datasets $X$ and $Y$, implies that $X = Y$.

In effect, the DSSIM quantization step is useful in terms of better differentiating changes in visual quality of the data in this region of interest where the data compression is quite conservative, meaning that differences between the two datasets are quite small and the SSIM and SSIM-like variants should be nearly 1.0.  Consider the situation when the two datasets are nearly equivalent
and the data is varying smoothly in a local patch, then quantizing the data values will amplify small differences in data that spans quantization bin boundaries and remove small differences within a quantization bin. Even though most of the differences will fall within a bin (and thus be zero after quantization), when the differences between bin boundaries are an order of magnitude or so larger than 
differences between the datasets (as happens when the DSSIM is nearly 1.0), then in our experience, the quantization step allows the DSSIM to highlight the differences in the data in a similar manner to transforming data to an image with 256 colors as done with the SSIM. While the exact DSSIM value depends on the number of quantization bins, of course, we find that the DSSIM with sensible choices for numbers of bins (e.g., 128, 256, 512, 1024) better differentiates  differences near 1.0 than the DSSIM without quantization.

\subsection{A cutoff threshold for DSSIM}

The primary reason for developing the DSSIM as a replacement to the SSIM is to reduce computational costs when evaluating compression effects on large amounts of data.  When using the SSIM to quantify the similarity of an image generated with the compressed climate data to that with the original, we use a so-called cutoff threshold, above which the quality of similarity is deemed acceptable. More specifically, in \cite{baker2019} we determined that the SSIM with a cutoff threshold of $0.99995$ indicated whether climate scientists would be able to detect a difference in CESM diagnostic images after compression.  This threshold, based on a large user study, is much tighter than the generally accepted SSIM indistinguishability threshold of $0.99$ (e.g., \cite{nilsson2020understanding}) and the $0.98$ suggested for medical imaging (e.g., \cite{g2012, wegener2010}). Therefore, 
note that because we are using the SSIM with a hard threshold, small effects from plotting choices can be important in regions near the threshold.  Thus, the DSSIM's 
independence from plotting decisions is a desirable addition to the cost-savings gained by not generating specific images.
However, to use the DSSIM instead of the SSIM for evaluating the climate data, we need to determine an appropriate cutoff threshold for the DSSIM, and conducting another large user study is not feasible at this time.
Instead, we use 
statistical techniques and the previous study results to show
that the DSSIM can be used to evaluate lossy compression artifacts in climate data with an appropriate cutoff threshold.

\begin{figure}[t]
\centering
\subfloat[][]{\includegraphics[width=6.5cm]{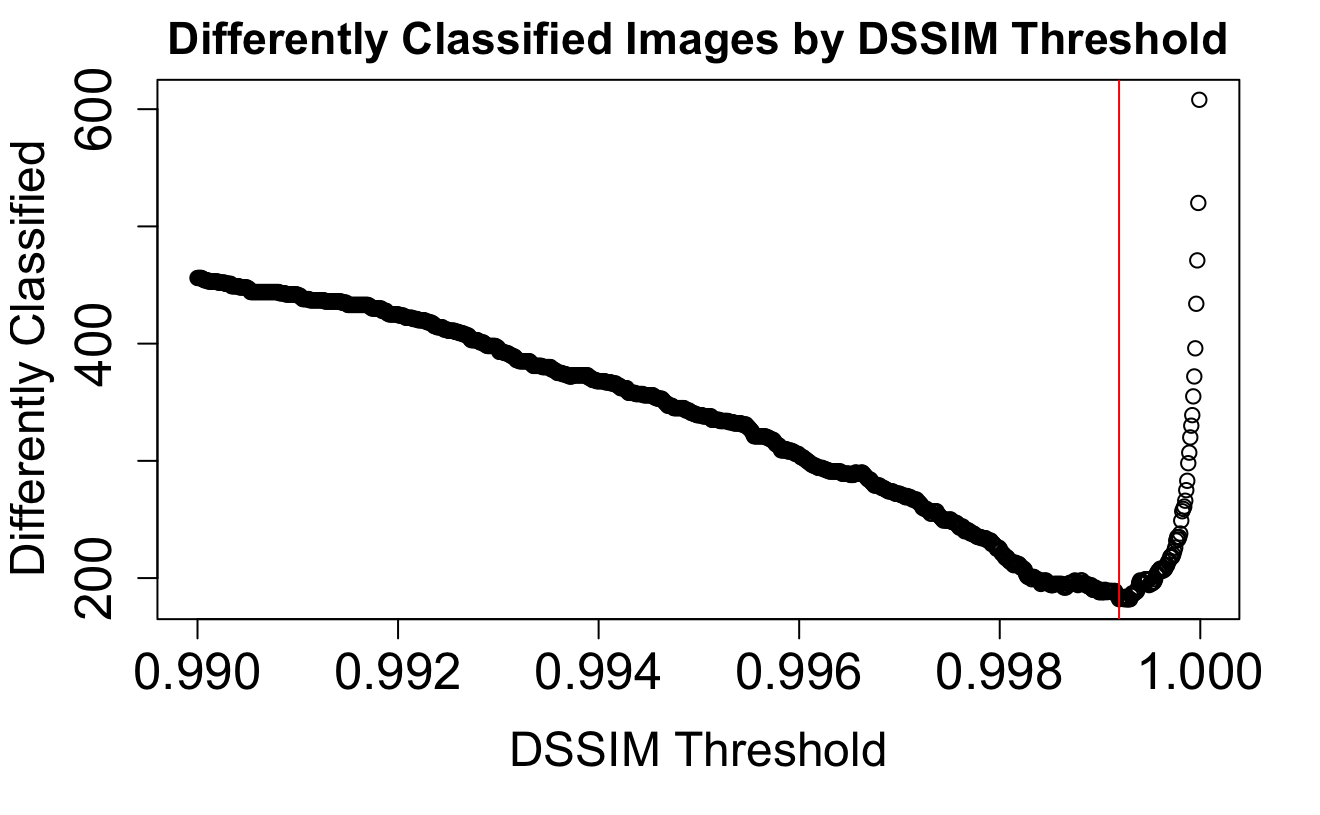} }
\newline
\subfloat[][]{\includegraphics[width=7cm]{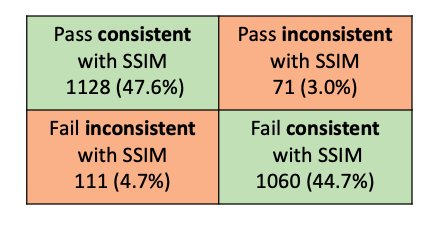} }
\caption{ The top plot
shows the number of differently classified (i.e., inconsistent) datasets by DSSIM threshold, which is minimized when the DSSIM threshold is 0.99919. The bottom plot contains the classification matrix for a DSSIM threshold of 0.99919.}
\label{f:thresh}
\end{figure}

Ideally, we want to find a DSSIM threshold such that datasets that pass the SSIM threshold test also pass the corresponding DSSIM threshold test, and datasets that fail the SSIM threshold test also fail the corresponding DSSIM threshold test. 
We define a compressed dataset to be a true ``pass" if the SSIM value meets or exceeds the 0.99995 threshold, otherwise we consider that dataset to be a true ``fail". 
One method to assess an appropriate DSSIM threshold 
is with classification matrices, which are commonly used tools to evaluate the results of a classification model. In our case, the classification matrix is a $2 \times 2$ matrix where the
columns correspond to the true pass or fail status of the data as determined by the SSIM.
The rows correspond to whether our model (i.e., the DSSIM) passes or fails the dataset, which is based on whether the DSSIM is above (pass) or below (fail) the DSSIM threshold being tested.
This setup means that the diagonal entries of the matrix correspond to the number of instances where there is agreement or consistency between the DSSIM and the SSIM, and the off-diagonal elements correspond to the number of instances where the DSSIM and SSIM disagree (an inconsistency) in their classification decision.
For example, if we take the first 3 time slices from the 79 2D monthly variables in the CESM-LENS dataset and apply the \textsc{zfp} compressor using  10 parameters for $p$ ($p = 6,8,10,12,14,16,18,20,22,24$), then we obtain 2370 SSIM values obtained by plotting and comparing the original dataset and the compressed dataset. 
As before, SSIM and DSSIM values are computed via \textsc{ldcpy} with the default settings.
In Figure \ref{f:thresh}, the top plot shows the number of images for which the DSSIM result was classified differently than the SSIM result (i.e., ``inconsistent") for a range of DSSIM thresholds. The bottom plot is the classification matrix corresponding to a DSSIM threshold of $0.99919$, which minimizes the inconsistent results (the sum of the off-diagonal entries in orange). 
Note that alternatively one could choose to minimize either the number of inconsistent fails or inconsistent passes, depending on the use case.

\begin{figure*}
\begin{center}
{\includegraphics[width=14.5cm]{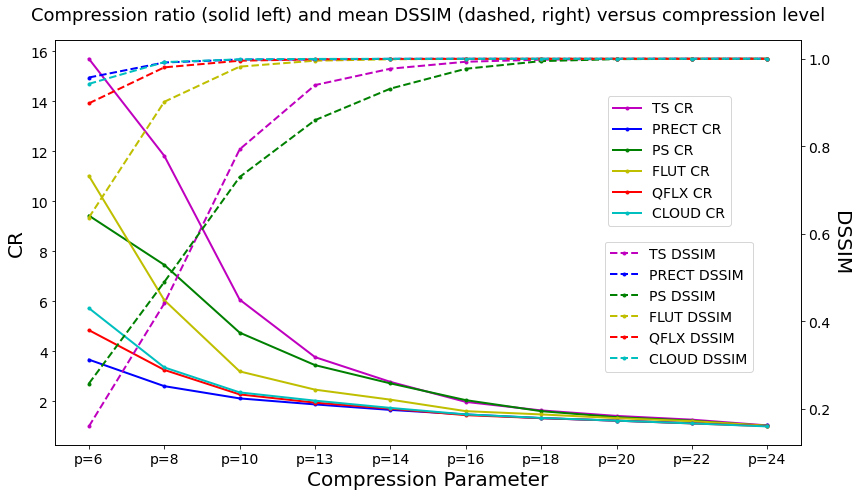} }
\caption{The relation between various amounts of lossy compression and the compression ratio (CR)  and DSSIM values for the variables in Table \ref{t:var_props}. The horizontal axis indicates the different fixed precision parameters for \textsc{zfp} compression from most (left) to least aggressive (right). The dashed lines correspond to the DSSIM values (right axis) and the solid lines correspond to the CR (left axis). The compression ratio is the losslessly compressed file size divided by the lossy compressed file size.}
\label{f:5var}
\end{center}
\vspace{-4mm}
\end{figure*}

This analysis for finding a cutoff threshold assumes that the data distribution in our analysis is representative of the data distribution in practice.  Because the CESM variables have quite different characteristics,
an appropriate DSSIM threshold based on only a single variable may be slightly different (lower or higher).  However, because the SSIM threshold determined in \cite{baker2019} is quite conservative, we find that the corresponding DSSIM
threshold of 0.99919 is conservative enough to use on all CESM variables. In practice, we often reduce this threshold further to allow for more aggressive data compression (e.g., to 0.995 or 0.95).  

Figure \ref{f:5var} gives an indication of how much compression can be achieved with \textsc{zfp} in fixed-precision mode for the variables in Table \ref{t:var_props}, including how the data reduction corresponds to the DSSIM.  The compression ratio (CR) is the size of the losslessly compressed data divided by the lossily compressed data with \textsc{zfp}.  
We see that the CR decreases as the \textsc{zfp} precision parameter $p$ increases, as expected.
CR values near 1.0 mean that the lossy compression is so conservative that there is little reduction beyond what lossless compression has achieved.
The DSSIMs increase in a nonlinear fashion, which is consistent with the idea that we get diminishing returns in the data fidelity as we approach lossless compression.  Moreover, there is clearly a limit to the amount of compression possible while maintaining a high DSSIM value.
The larger the CR, the lower the DSSIM tends to be, which can easily be seen by noting how the colors of
the dashed and solid lines are nearly in reverse order from top to bottom. 
Further
examinations of DSSIM in the context of data compression for climate model data can be found in technical reports  \cite{pinard2021} and \cite{pinard2021-behavior}.

\subsection{SSIM vs. DSSIM speedup }
Finally, we show that, as expected, computing the DSSIM is much faster than computing the SSIM.
   We time the calculations of the DSSIM and the SSIM using the implementations in the \textsc{ldcpy} package via a Jupyter notebook.  We again use the TS and PRECT variables from the CESM-LENS data that are included with \textsc{ldcpy}.
Timing results  (in seconds) are given in Table \ref{t:times}.  The algorithmic cost of the DSSIM and SSIM are similar, so the difference in time is due to the rendering of the images from the floating-point data needed to calculate the SSIM. While the actual times for these calculations will vary depending on the computing platform (these were performed on a laptop), the speedup indicates what type of performance can be gained by using the DSSIM instead of the SSIM. While the rendering performance could of course be improved by using different hardware (e.g., GPUs) or more optimized software, avoiding the rendering altogether will always be less expensive. Whether this savings is significant enough to matter to the user will be specific to their particular hardware, software stack, and data volume.
For our application, the DSSIM is more competitive from a computational standpoint to other metrics for data similarity applied to the floating-point data, like the mean-squared error (MSE) or the peak signal-to-noise ratio (PSNR). (Note that
the MSE and PSNR can be applied to either floating-point or  image data. When applied to floating-point data, it is well known they are not particularly indicative of visual similarity. However, they may be useful for evaluating visual similarity in some cases when applied to image data.)

The time savings for the DSSIM are important for comparing compressor results with climate data, particularly as we automate testing and must evaluate the data quality and similarity on large amounts of data.
Indeed, in previous work we had found the SSIM to be a better indication of image similarity than other measurements, but its relative expense made its use hard to justify. The cost of the DSSIM, on the other hand, is more reasonable for our large-scale evaluations and has proven useful in practice.

\begin{table}[tb]
\centering
\caption{Timings for computing the SSIM and DSSIM via the \texttt{DataCalcs} object in \textsc{ldcpy}. Times reported are the fastest of 5 executions in seconds. The first time slice is used for each variable. } \label{t:times}
\footnotesize
\begin{tabular}{|l|l|l|l|}
\hline
variable  & SSIM (s) & DSSIM (s) & speedup \T \B \\
\hline \hline
TS  & 7.72 & .0396 & 145x \T \\
PRECT  & 7.89 & .0373 & 212x \B \\
\hline
\hline
\end{tabular}
\vspace{-4mm}
\end{table}

\section{Concluding Remarks}\label{sec:concl}

In this manuscript, we have proposed an alternative to the popular SSIM that can be applied directly to floating-point data.  
Applying the DSSIM to the floating-point data is computationally cheaper than generating images from the data and then applying the SSIM.   
This reduced computational cost is quite important when analyzing large volumes of data in an automated fashion and is appropriate for situations in which we need only a general idea of whether images created from the data will be similar. 
An additional benefit is that the DSSIM is independent of plot-specific choices that can affect the SSIM.
The DSSIM is implemented in the \textsc{ldcpy} (Large Data Comparison for Python)  package \cite{pinard20}.

 While conceptually simple, the DSSIM has been tremendously beneficial to us for comparing lossily compressed to uncompressed climate model data. Prior to our development of the DSSIM, 
the SSIM had become an important measurement in our compression evaluation toolkit, largely due to its intuitiveness for the user and ability to represent visual similarity well.
However, the SSIM was prohibitively more expensive to compute than all other measurements in the toolkit on large data volumes because of the image generation requirement, motivating us to  propose the DSSIM.
In practice, we now have been using the DSSIM instead of the SSIM
with success in terms of identifying compression artifacts and saving compute time when evaluating data compression.
While we have only evaluated the DSSIM in the context of comparing climate model simulation data,
we are optimistic that it could be a useful measurement in other application areas as well - especially those producing large volumes of simulation data.

\section*{Acknowledgment}
We acknowledge high-performance computing support from Cheyenne (doi:10.5065/D6RX99HX) provided by NCAR's Computational and Information Systems Laboratory, sponsored by the National Science Foundation.

\bibliographystyle{abbrv-doi}

\bibliography{refs}
\end{document}